\documentclass[10pt,conference]{IEEEtran}
\IEEEoverridecommandlockouts

\usepackage{cite}
\usepackage{amsmath,amssymb,amsfonts}
\usepackage{graphicx}
\usepackage{textcomp}
\usepackage{xcolor}
\usepackage{todonotes} 

\usepackage{multicol}
\usepackage{hyperref}
\usepackage{psfrag}
\usepackage{epstopdf}
\usepackage{amsfonts}
\usepackage{mathrsfs}
\usepackage{pifont}
\usepackage{verbatim}
\usepackage{upgreek}
\usepackage{color}
\usepackage{algorithm}
\usepackage{mdwmath}
\usepackage{mdwtab}
\usepackage{amssymb,amsmath}
\usepackage{amsmath}
\usepackage{amsthm}
\usepackage{epstopdf}
\usepackage{cite}
\usepackage{tikz}
\usepackage{scalefnt}
\usepackage{subfigure}
\usepackage{MnSymbol}
\usepackage{setspace}
\usepackage{mathtools}
\usepackage{psfrag}
\usepackage{balance}
\usepackage{soul}
\usepackage[font=scriptsize]{caption}
\usepackage{algpseudocode}

\usepackage{multirow}
\usepackage{booktabs}

\usepackage{cases}

\label{English Chars}

\newcommand{\bC}{\textbf{C}}

\newcommand{\bO}{\textbf{O}}

\newcommand{\bW}{\textbf{W}}

\newcommand{\bX}{\textbf{X}}

\newcommand{\bY}{\textbf{Y}}

\newcommand{\bZ}{\textbf{Z}}

\def\BibTeX{{\rm B\kern-.05em{\sc i\kern-.025em b}\kern-.08em
    T\kern-.1667em\lower.7ex\hbox{E}\kern-.125emX}}
\begin{document}

\title{An Optimization Driven Link SINR Assurance
in RIS-assisted Indoor Networks}
\author{\IEEEauthorblockN{Cao Vien Phung \IEEEauthorrefmark{3}, Max Franke \IEEEauthorrefmark{2}, Ehsan Tohidi \IEEEauthorrefmark{1}\IEEEauthorrefmark{2}, June Heinemann \IEEEauthorrefmark{3}, \\ Andr\'e Drummond \IEEEauthorrefmark{3}, Stefan Schmid \IEEEauthorrefmark{2}, Slawomir Sta\'nczak \IEEEauthorrefmark{1}\IEEEauthorrefmark{2}, and Admela Jukan \IEEEauthorrefmark{3}}
\IEEEauthorblockA{\textit{\IEEEauthorrefmark{3} Technische Universit\"at Braunschweig, Germany} \\\textit{\IEEEauthorrefmark{1} Fraunhofer Institute for Telecommunications, Heinrich-Hertz-Institut, Berlin, Germany}  \\ \textit{\IEEEauthorrefmark{2} Technische Universit\"at Berlin, Germany} \\
Email: (c.phung, andre.drummond, a.jukan, j.heinemann)@tu-bs.de, (mfranke, stefan.schmid)@tu-berlin.de, \\ (ehsan.tohidi, slawomir.stanczak)@hhi.fraunhofer.de}
}
\maketitle

\begin{abstract}
Future smart factories are expected to deploy applications over high-performance indoor wireless channels in the millimeter-wave (mmWave) bands, which on the other hand are susceptible to high path losses and Line-of-Sight (LoS) blockages. Low-cost Reconfigurable Intelligent Surfaces (RISs) can provide great opportunities in such scenarios, due to its ability to alleviate LoS link blockages. In this paper, we formulate a combinatorial optimization problem, solved with Integer Linear Programming (ILP) to optimally maintain connectivity by solving the problem of allocating RIS to robots in a wireless indoor network. Our model exploits the characteristic of nulling interference from RISs by tuning RIS reflection coefficients. We further consider Quality-of-Service (QoS) at receivers in terms of Signal-to-Interference-plus-Noise Ratio (SINR) and connection outages due to insufficient transmission quality service. Numerical results for optimal solutions and heuristics show the benefits of optimally deploying RISs by providing continuous connectivity through SINR, which significantly reduces outages due to link quality. 
\end{abstract}
\vspace{-0.34cm}
\begin{IEEEkeywords}
Interference, millimeter-wave (mmWave), Reconfigurable Intelligent Surface (RIS), reliability, smart factory.
\end{IEEEkeywords}

\section{Introduction} \label{intro}
Reconfigurable Intelligent Surfaces (RISs) in upcoming 6G networks have been widely projected as breakthrough technology in wireless networks, especially in mmWave and Sub-THz links, to alleviate problems of Line of Sight (LoS) links between base stations and receivers. Currently, applications such as 6G campus networks in smart factories are being tested and deployed, along with other widely investigating 6G mesh wireless networks, such as in vehicular and health care domains.  A smart factory scenario is especially interesting as it typically envisions dynamic industrial robots connected to static Base Stations (BS), or RISs. 
Each RIS can accommodate multiple robots, whereby attention needs to be paid to the resulting quality of transmission in wireless links.

 Despite being a highly controlled environment, robots can experience wireless link outages due to mobility and interference, which is critical, especially if the outage time exceeds a threshold allowed by each robot in factory. There are issues of Signal-to-Interference-plus-Noise Ratio (SINR), due to competing and interfering robots that simultaneously connect to the same BS and RISs. The problem of optimal RIS allocation for multiple robots, and the related co-channel interference-avoiding strategies in smart factories are new and unsolved, however critical to the emerging applications envisioned in 6G networks. We study the optimization driven problem of SINR based link quality, and propose in this paper an ILP and heuristics to optimize the time that each robot is served by the network under the transmission quality constraints. We show that optimizations can minimize link outages and reduce robot service failures. The performance results also quantify the benefits of exploiting RISs as compared to the smart factories with no RISs.

The remainder of the paper is organized as follows. Section \ref{relatedwork} presents related work. Section \ref{systemmodel} presents the system model. Section \ref{ILPpro} formulates the optimization problem. Section \ref{perevalu} shows performance evaluation.  Section \ref{conclu} concludes the paper.

\section{Related work} \label{relatedwork}
The geometric shape models of conical transmission beams from \cite{7820226,9822386} are used in this paper to identify areas causing interference. Further  studies \cite{10569578,whsglb} build an interference model based on such geometric analyses, which we also use as basis for interference computations in this study. Some geometric analyses, such as beam footprint in mmWave, are discussed in \cite{9386246} and are used here to identify interfering areas. Papers \cite{8796365,9840504} discuss the channel model for RIS-assisted networks, which is applied to this work. Moreover, paper \cite{10.1093/comjnl/bxaa083} presents the interference model for indoor THz communications in which any receiver suffers interference from any users, which is also used in this paper. This paper extends our previous work \cite{10539144}, which focuses on single connection reliability, whereas this paper focuses on multiple access from robots to RIS. The novelty of this paper is that it integrates the QoS with SINR, as well as the characteristic of nulling interference at RISs into the ILP. Regarding the interference nulling solution at each RIS, paper \cite{9681803} proposes an alternating projection algorithm that allows $U$ robots to be simultaneously served without interference, which is our goal as well.

\section{System model} \label{systemmodel}
\subsection{Reference scenario} \label{refsce}
\begin{figure}[t]
    \centering
     \includegraphics[width=0.5\textwidth]{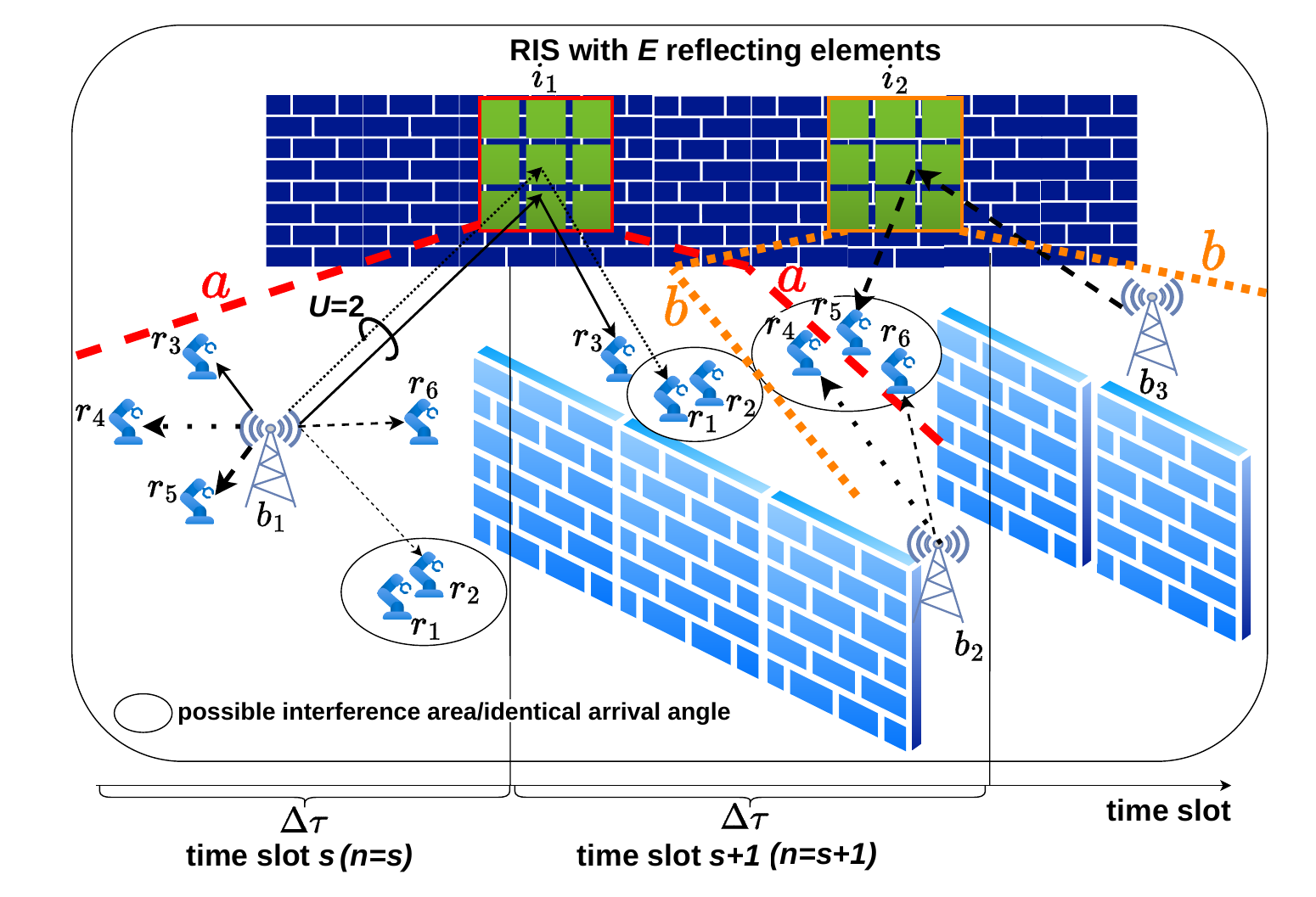}
        \caption{The reference scenario of a RIS-assisted smart factory.}
        \label{scenario}
        
\end{figure}

The reference scenario of smart factory is shown in Fig. \ref{scenario}. We show a set $B=\{b_1,b_2,b_3\}$ of $3$ static Base Stations (BSs),  a set $I=\{i_1,i_2\}$ of $2$ static RISs, and a set $R=\{r_1,r_2,r_3,r_4,r_5,r_6\}$ of $6$ dynamic robots. Each robot connects to one BS via either a LoS mmWave channel or a virtual LoS channel facilitated by a RIS if direct LoS links between the BSs and the robot are obstructed. As robots move, the LoS or virtual LoS links change over time, with the time interval of interest $[0,\tau]$ divided into a set $N=\{n=s,n=s+1\}$ of $2$ time slots in this example. Each time slot lasts for a duration of $\Delta \tau$; thus, $\tau = |N| \Delta \tau $. 


Let us assume that channels are deterministic and known in advance for every location of the robots at any time slot. RISs are mounted on fix locations (e.g., walls) with LoS to BSs. We define the coverage regions for each RIS.  In Fig. \ref{scenario},  RIS $i_1$ and RIS $i_2$ have the coverage regions limited by the lines $a$ and $b$, respectively. During time slot $n=s+1$, RIS $i_1$ can cover robots $\{r_1,r_2,r_3,r_4\}$, whereas RIS $i_2$ can cover robots $\{r_4,r_5,r_6\}$. Robot $r_4$ can be covered by both RIS $i_1$ and RIS $i_2$. Moreover, assume BSs have coverage regions with all devices in their LoS links. In Fig. \ref{scenario}, during $n=s$, BS $b_1$ covers all robots $\{r_1,r_2,r_3,r_4,r_5,r_6\}$ and RIS $i_1$, whereas these robots moving with certain and trajectories known \emph{a priori}, and that all RISs $\{i_1,i_2 \}$ can be covered by BS $b_2$. If each robot is neither covered by BSs nor by RISs, we consider this a link outage in a time slot measured. Also, a link can experience outage if its SINR is lower than a threshold $\Psi_r$. If a maximum number of times that a robot can accept consecutive outages is $K_r$ time slots, before we define a service failure for any consecutive outage periods higher or equal to $K_r$.

Each RIS has $E$ reflecting elements, whereby, based on \cite{9681803}, if $E$ elements are slightly larger than a threshold of $2U(U-1)$, then $U$ communication pairs between BSs and robots can be simultaneously transmitted without interference, by adjusting RIS reflection coefficients. However, to eliminate interference, the arrival angles of signals from robots must be different \cite{9681803}. If arrival angles overlap, interference nulling is not possible. In Fig. \ref{scenario}, assume that each RIS can accommodate a maximum of $U=2$ communication pairs simultaneously with nulling interference. Robots $r_1$ and $r_2$ in the circle area during time $n=s+1$ have the same arrival angle, and cannot be simultaneously scheduled through RIS $i_1$. Assume that only robots $r_1$ and $r_3$ can be chosen to access simultaneously via RIS $i_1$ with nulling interference. The RIS configuration from covering one robot to another takes $D$ time slots. If a RIS is available, it can immediately serve the robots, i.e., without reconfiguration delay. BSs on the other time do not induce configuration delay in their corresponding LoS coverage regions.

\subsection{Allocation strategy} \label{allostra}
Based on the reference scenario, we now illustrate how the optimal allocation of BSs and RISs to robots can minimize the outage. In Fig. \ref{scenario}, if an arrow exists between one robot and a BS/RIS, that BS/RIS is allocated to that robot. We now analyze the allocation strategy in Fig. \ref{scenario}. During $n=s$,  two robots $r_1$ and $r_2$ belonging to the circle area may interfere. BS $b_1$ simultaneously transmits its signal to these two robots if the SINR of robots $r_1$ and $r_2$ is larger than or equal to the threshold $\Psi_{r_1}$ and $\Psi_{r_2}$, respectively. Otherwise, only one of the robots, e.g., $r_2$, receives the signal from BS $b_1$ during $n=s$. During time $n=s+1$, considering the coverage region of RIS $i_2$, it can cover three robots $\{r_4,r_5,r_6 \}$. As these three robots have the same arrival angle, only one of them, e.g., $r_5$ in Fig. \ref{scenario}, is chosen to be allocated by RIS $i_2$. The remaining two robots $r_4$ and $r_6$ are assumed to be simultaneously scheduled with robot $r_5$ via BS $b_2$, if the interference among them still satisfies the SINR of these three robots $r_4$, $r_5$, and $r_6$ larger than or equal to the threshold $\Psi_{r_4}$, $\Psi_{r_5}$, and $\Psi_{r_6}$, respectively.

At the same time slot $n=s+1$, and considering the coverage region of RIS $i_1$,   robots $\{ r_1,r_2,r_3,r_4\}$ can be served. As robots $r_1$ and $r_2$ have the same arrival angle, they cannot be scheduled simultaneously via RIS $i_1$. As only $U=2$ robots are simultaneously served via each RIS, there are five options for RIS $i_1$: $\{r_3,r_1\}$, $\{r_3,r_2\}$, $\{r_3,r_4\}$, $\{r_1,r_4\}$, and $\{r_2,r_4\}$ for simultaneously connecting via RIS $i_1$. However, since $r_4$ can be directly connected to BS $b_2$, the options including this robot can be discarded. From the two options left, $\{r_3,r_1\}$ and $\{r_3,r_2\}$, $\{r_3,r_1\}$ is the best choice because the allocation history of $r_2$ was already connected to BS $b_1$ during $n=s$. In this way, we can mitigate the service failures in which each robot expects the outage in less than $K_r$ consecutive time slots. If BS $b_2$ is allocated to robot $r_2$ during $n=s+1$, the SINR of robot $r_1$ may become lower than the threshold $\Psi_{r_1}$. Therefore, robot $r_2$ experience an outage during $n=s+1$.

Finally, due to the characteristic of nulling interference \cite{9681803}, the interference never occurs for the robots accessing the same RIS. However, those robots still suffer interference from the robots allocated by other RISs and the BSs with LoS transmission. For example, if robot $r_4$ is assumed to be allocated by RIS $i_1$ during $n=s+1$ in Fig. \ref{scenario}, it will not get the interference from robots $r_1$, $r_2$, and $r_3$. However, it still suffers interference from the received signal of robot $r_5$ via RIS $i_2$ and from the received signal of robot $r_6$ via the LoS transmission of BS $b_2$. This can be explained by the fact that the solution to the RIS interference nulling problem is only valid for the robots with access to the RIS itself.

\subsection{Analytical model of SINR} \label{interanmodel}
We now calculate SINR for each robot $r \in R$ with consideration of path loss in mmWave networks. The SINR for each robot $r$ at time slot $n$, connected via RIS $i \in I$ is:
\begin{equation} \label{SNIRirn}
 S_{i,r,n} = \frac{P_{i,r,n} G_b G_r}{P_o+ \sum_{r\prime \in R \setminus r} \left(\sum_{b \in B} \xi_{b,r^\prime r,n} \right) + \left( \sum_{i^\prime\in I \setminus i}  \xi_{i^\prime,r^\prime r,n} \right)}
\end{equation}

The transmitting antenna gain $G_b$ of the BS $b$ is given in \cite{7820226}:
\begin{equation}\label{G_b_BS}
    G_b = \frac{2}{1-cos\left( \frac{\theta}{2} \right)},
\end{equation}
whereas $\theta$ is the antenna directivity angle. $G_r$ is the receiving antenna gain of robot $r$, whereas $G_b=G_r$. Based on \cite{8796365,10569578,9840504}, with optimal phase, the signal of each robot $r$ obtained via RIS $i$ at each time slot $n$ is:
\begin{equation} \label{Pirn}
  P_{i,r,n} =   P_b|H_{(b,i,n)} \cdot  E \cdot H_{(i,r,n)} |^2,
\end{equation}
whereas $P_b$ is the transmitting power of BS $b \in B$, $E$ is the number of RIS elements, the channel transfer functions $H_{(b,i,n)}$ between BS $b$ and RIS $i$ at the time slot $n$ and $H_{(i,r,n)}$ between RIS $i$ and robot $r$ at the time slot $n$ is:
\begin{equation} \label{Hcommon}
H = \left( \frac{c}{4\pi fd} \right),
\end{equation}
whereas \eqref{Hcommon} can be replaced by $H_{(b,i,n)}$ or $H_{(i,r,n)}$, $c$ is the light speed, $f$ is the mmW frequency, and $d$ is the distance; $d$ can be replaced by $d_{b,i,n}$, i.e., distance between BS $b$ and RIS $i$, or by $d_{i,r,n}$, i.e., between RIS $i$ and robot $r$, at the time slot $n$). Also, in \eqref{SNIRirn}, the noise power $P_o = k T V$, where $k$ is the Boltzmann constant, $T$ is the absolute temperature in Kelvin, and $V$ denotes the bandwidth. The undesired interference $\xi_{b,r^\prime r,n}$ from the LoS transmission between BS $b$ and robot $r^\prime$ causing on robot $r$ at the time $n$ is:
\begin{equation} \label{interference_delta}
   \xi_{b,r^\prime r,n} = \delta_{b,r,n} G_b G_r, 
\end{equation}
where the undesired LoS interference $\delta_{b,r,n}$ without the antenna gain obtained at robot $r$ from BS $b$ at the time $n$ is:
\begin{equation} \label{inter_withoutgain}
    \delta_{b,r,n} = P_b |H_{(b,r,n)}|^2,
\end{equation}
where \eqref{Hcommon} is replaced by $H_{(b,r,n)}$, and $d$ is replaced by $d_{b,r,n}$, i.e., distance between BS $b$ and robot $r$ at the time $n$. The undesired interference $\xi_{i^\prime,r^\prime r,n}$ from  the virtual LoS link with RIS $i^\prime$ for robot $r^\prime$ causing on robot $r$ at the time $n$ is:
\begin{equation} \label{inter_virtual}
  \xi_{i^\prime,r^\prime r,n} =  \delta_{i^\prime,r,n} G_b G_r, 
\end{equation}
where the undesired virtual LoS interference $\delta_{i^\prime,r,n}$ without the antenna gain obtained at robot $r$ via RIS $i^\prime$ at the time slot $n$ is applied as \eqref{Pirn}. As discussed in section \ref{refsce}, since the robots connected to the same RIS will not get the interference due to either transmission scheduling at different times or interference removed
by the solution proposed in \cite{9681803}, we do not consider the interference from those robots in \eqref{SNIRirn}. Finally, the SINR for robot $r$ at the time $n$ if directly connected to BS $b$ is:
\begin{equation} \label{SNIRbrn}
 S_{b,r,n} = \frac{P_{b,r,n} G_b G_r}{P_o+ \sum_{r\prime \in R \setminus r} \left(\sum_{b \in B} \xi_{b,r^\prime r,n} \right) + \left( \sum_{i\in I}  \xi_{i,r^\prime r,n} \right) },
\end{equation}
where the signal power $P_{b,r,n}$ of robot $r$ directly obtained from BS $b$ at the time $n$ is applied as \eqref{inter_withoutgain}, and the interference $\xi_{b,r^\prime r,n}$ and $\xi_{i,r^\prime r,n}$ are applied as \eqref{interference_delta} and \eqref{inter_virtual}, respectively.

\section{Problem formulation} \label{ILPpro}
We now formulate the optimization problem and build an ILP program that minimizes the connection outages for the reference scenario described. The input parameters and binary variables are summarized in Table \ref{tab:table2}. The minimization objective is achieved by generating the optimal allocation for robot $r\in R$ on one of the BSs in $B$, or RISs in $I$. The objective function is defined as the following: 
\begin{equation}\label{obfun}
     \underset{\bW, \bX, \bO, \bY, \bC, \bZ}{\min} \sum_{r \in R}\sum_{n \in N} O_{r,n},
\end{equation}
subject to the following constraints:
\begin{equation} \label{cons2}
   \sum_{b\in B} X_{b,r,n} + \sum_{i \in I} X_{i,r,n}  \leqslant 1, \; \forall r \in R,n \in N,
\end{equation}
where constraint \eqref{cons2} is for each robot $r$ at $n$, that robot connects to only one BS $b$ / RIS $i$ via the binary variables $X_{b,r,n}/X_{i,r,n}$,
\begin{equation} \label{cons3}
    \sum_{r \in m} X_{i,r,n}  \leqslant 1, \; \forall i \in I, m \in M_{i,n},n \in N,
\end{equation}
where constraint \eqref{cons3} is for RIS $i$ at time slot $n$, there exists a set $M_{i,n}$. Each $m \in M_{i,n}$ is a set of robots with the variables $X_{i,r,n}$ that their receiving beams are overlapped together, resulting in conflicts. With the interference nulling solution \cite{9681803}, the robots in $m$ cannot be simultaneously scheduled, i.e., 
\begin{equation} \label{cons4}
    \sum_{r \in R} X_{i,r,n}  \leqslant U, \; \forall i \in I,n \in N,
\end{equation}
where constraint \eqref{cons4} is for RIS $i$ at $n$, $U$ robots with variable $X_{i,r,n}$ are simultaneously scheduled, see Section \ref{systemmodel},
\begin{subnumcases}{}
  \frac{ X_{b,r,n }  P_{b,r,n}  G_b  G_r  +  \omega }{P_{o}+ \sum_{r^\prime \in R \setminus r } (\varpi  +   \vartheta)}  \geqslant \Psi_r,\forall r \in R, b \in B,n \in N, \;\;\;\;\;\;  \label{cons5} \\
 X_{b,r,n}+Z_{b,r,n}  = 1, \; \forall r \in R, b \in B,n \in N, \label{add1} 
\end{subnumcases}
where constraint \eqref{cons5} is SINR of each robot link for $r$ at $n$ connected to BS $b$ with variable $X_{b,r,n}$, whereas \eqref{cons5} can be referred to \eqref{SNIRbrn}, $\omega=\mu Z_{b,r,n} $, $ \varpi=\sum_{b \in B} X_{b,r^\prime,n}  \xi_{b,r^\prime r,n}$, and $\vartheta=\sum_{i\in I} X_{i,r^\prime,n}  \xi_{i,r^\prime r,n}$, and SINR $\geq \Psi_r$ (threshold). We define a dummy binary variable $Z_{b,r,n}$ with a constant $\mu$ assigned large enough, where with \eqref{add1}, if $X_{b,r,n}=1$, $Z_{b,r,n}=0$; else $Z_{b,r,n}=1$. With $Z_{b,r,n}$, \eqref{cons5} is always satisfied, if $X_{b,r,n}=0$,
\begin{subnumcases}{}
  \frac{ X_{i,r,n}  P_{i,r,n}  G_b G_r + \omega^{\prime} }{P_o + \sum_{r^\prime \in R \setminus r }  (\varpi +  \vartheta^\prime )   }  \geqslant \Psi_{r}, \forall r \in R,i \in I,n \in N, \;\;\;\;\;\; \label{cons6} \\
 X_{i,r,n}+Z_{i,r,n}  = 1, \; \forall r \in R, i \in I,n \in N, \label{add2}  
\end{subnumcases}
where constraint \eqref{cons6} can be explained as \eqref{cons5}, but robot $r$ is connected to RIS $i$ with variable $X_{i,r,n}$, whereas  \eqref{cons6} can be referred \eqref{SNIRirn}, $\omega^{\prime}=\mu Z_{i,r,n}$, and $\vartheta^\prime=\sum_{i^\prime \in I \setminus i} X_{i^\prime, r^\prime,n} \xi_{i^\prime, r^\prime r,n}$. The binary variable $Z_{i,r,n}$ in \eqref{cons6} can be referred to $Z_{b,r,n}$ in \eqref{cons5}, where with \eqref{add2}, if $X_{i,r,n}=1$, $Z_{i,r,n}=0$; else $Z_{i,r,n}=1$,
\begin{equation} \label{cons7}
    \frac{1}{D}\sum_{\bar{n}=\max\{n-D+1,0\}}^n X_{i,r,\bar{n}}  \leqslant Y_{i,r,n}, \; \forall r \in R, i \in I,n \in N,
\end{equation}
where constraint \eqref{cons7} states the allocation history of RISs during the last $D$ time slots. If RIS $i$ is allocated to robot $r$ at least once during time $[n-D+1,n]$, the binary variable $Y_{i,r,n}=1$. Otherwise, $Y_{i,r,n}=0$,
\begin{equation} \label{cons8}
    \left( \frac{1}{|R|}\sum_{r \in R} Y_{i,r,n} \right) -\frac{U}{|R|} \leqslant C_{i,n}, ~~ \forall i \in I,n \in N ,
\end{equation}
where constraint \eqref{cons8} considers the required time of reconfiguration. If the binary variable $C_{i,n}=1$, RIS $i$ reallocated during $D$ time slots is not ready. Otherwise, $C_{i,n}=0$. With \eqref{cons7}, $C_{i,n}=1$, if $Y_{i,r,n}=1$ at least $U+1$ different robots $r$,
\begin{equation} \label{cons9} 
 \sum_{\bar{n}=\max\{n-K_r+1,0\}}^n O_{r,\bar{n}} < K_r, \; \forall r \in R,n \in N   ,
\end{equation}
where \eqref{cons9} considers the connection status of the robots with the binary variable $O_{r,n}$. If $O_{r,n}=1$, then the robot $r$ gets an outage at the time slot $n$. Otherwise, $O_{r,n}=0$. This constraint defines a service failure for each robot $r$ with $K_r$ time slots. The summation of the outage variable $O_{r,n}$ of each robot $r$ over $K_r$ consecutive time slots should be less than $K_r$,
\begin{subnumcases}{}
  O_{r,n}+\sum_{i \in I} W_{i,r,n}+\sum_{b \in B} X_{b,r,n}  \geq 1, \forall r \in R,n \in N,  \label{cons10} \\
 W_{i,r,n}  \leqslant X_{i,r,n}, \forall i \in I,r \in R,n \in N,  \label{cons11} \\
 W_{i,r,n} \leqslant 1- C_{i,n}, \forall i \in I,r \in R,n \in N,  \label{cons12}
\end{subnumcases}
where \eqref{cons10} states that if each robot $r \in R$ at time slot $n$ is directly connected to BS $b$ via the variable $X_{b,r,n}$ or is allocated to the RIS $i$ in case of that available RIS $i$ via the binary variable $W_{i,r,n}$, which its constraints are presented in \eqref{cons11} and \eqref{cons12}. If $W_{i,r,n}=1$, each robot $r$ at each time slot $n$ is allocated to the ready RIS $i$. Otherwise, $W_{i,r,n}=0$.
The complexity of the ILP solution is given by: \begin{equation}\label{unitscal}
\begin{split}
O&(|I|\cdot |R| \cdot |N| \cdot [|R|(|B|+|I|)+D]+|I| \cdot |N| \cdot |M_w| \cdot |m_w|+ \\ & |R|\cdot |B| \cdot |N| \cdot[|R|(|B|+|I|)]+|R| \cdot |N| \cdot [|B|+|I|+K_r]),
\end{split}
\end{equation}
 With \eqref{cons2}, we need $|N| \cdot |R|(|B|+|I|)$ steps to search allocations for robots. There are $|I| \cdot |N| \cdot |M_w| \cdot |m_w|$ steps to find transmission scheduling solutions for RIS interference nulling in \eqref{cons3}, where $m_w \in M_w$ is the worst case, represented for $m \in M_{i,n}$ in \eqref{cons3}. We need $|I| \cdot |R| \cdot |N|$ steps to choose a maximum of $U$ robots simultaneously served through each RIS in \eqref{cons4}. We need $|R| \cdot |B| \cdot |N|\cdot [|R|(|B|+|I|)]$ steps to consider SINR threshold in \eqref{cons5} and $|R| \cdot |B| \cdot |N|$ steps for dummy variables in \eqref{add1}. We need $|R| \cdot |I| \cdot |N|\cdot [|R|(|B|+|I|)]$ steps to consider  SINR threshold in \eqref{cons6} and $|R| \cdot |I| \cdot |N|$ steps for dummy variables in \eqref{add2}. The RIS allocation history in \eqref{cons7} needs $|R| \cdot |I| \cdot |N| \cdot D$ steps. In \eqref{cons8}, we need $|I| \cdot |N| \cdot |R|$ steps to identify the status of RISs.  The service failure of robots is considered in \eqref{cons9} with $|R| \cdot |N| \cdot K_r$ steps. There are $|R| \cdot |N| \cdot (|B|+|I|)$ steps in \eqref{cons10}. Finally, we need $|I| \cdot |R| \cdot |N|$ steps for both \eqref{cons11} and \eqref{cons12}.

\begin{table}[t!]
  \footnotesize
  \centering
    \vspace{0.2 cm}
  \caption{List of input parameters and binary variables of ILP program.}
  \label{tab:table2}
  \begin{tabular}{ll}
    \toprule
 \textbf{Input} & \textbf{Meaning}\\
    \midrule
    \footnotesize
    $R,N$ & Set of robots and time slots, respectively. \\
    $B,I$ & Set of BSs and RISs, respectively.\\
    $M_{i,n}$ & Set of subsets $m$ with robots conflicted \\ 
              & at RIS $i$ at time $n$.\\
    $U$ & Robots simultaneously allocated at RIS.\\
    $\Psi_r$ & SINR threshold for robot $r$.\\
    $\xi$ & Interference on robot $r$ at $n$ caused by\\
    & comm. between BS $b$ / RIS $i$ and robot $r^\prime$, \\
    & $\xi=\{\xi_{b,r\prime r,n},\xi_{i,r\prime r,n}\}$.\\
    $P$ & Power with no gain obtained at robot $r$\\ 
                &  via BS $b$ / RIS $i$ at $n$, $P=\{P_{b,r,n},P_{i,r,n}\}$.\\
    $G$ & Antenna gain of BS $b$ / robot $r$, $G=\{ G_b,G_r\}$.\\
    $P_o$ & Noise power.\\
    $D$ & RIS configuration time.\\
    $K_r$ & Outage time slot threshold for robot $r$.\\
    $\mu$ & Large constant, e.g., $\mu=10^9$.\\
\toprule
  \textbf{Variables} & \textbf{Meaning}\\
  \midrule
  $X$ & Allocation of BS $b$ / RIS $i$ to robot $r$ at $n$, \\
  & $X=0/1$: unallocated/allocated, \\
  & $X=\{X_{b,r,n},X_{i,r,n}\}$.\\
$W_{i,r,n}$ & Allocation of RIS $i$ to robot $r$ at time $n$\\
             & as well as availability of RIS $i$ at time $n$, \\
             & $W_{i,r,n}=0/1$: unallocated/allocated.\\
 $O_{r,n}$ & Connection status of robot $r$ at time $n$,\\
           & $O_{r,n}=0/1$: connection/outage.\\
$Y_{i,r,n}$ & Allocation history of RIS $i$ to robot $r$ at $n$,\\
            & $Y_{i,r,n}=0/1$: unallocated/allocated   \\ 
            & at least once during $[n-D+1,n]$.\\           
$C_{i,n}$ & Availability status of RIS $i$ at time $n$,\\
          & $C_{i,n}=0/1$ : available/unavailable.\\
$Z$ & Dummy variable: $Z=\{Z_{b,r,n},Z_{i,r,n}\}=0/1$.\\
\bottomrule
  \end{tabular}
\end{table}

\section{Performance evaluation} \label{perevalu}
We now analyze the numerical results obtained by optimizations and heuristics. The heuristic connects each robot to one BS/RIS with the shortest path, based on \eqref{cons2}. Based on \eqref{cons4}, the heuristic selects randomly a maximum of $U$ robots simultaneously served by that RIS. Also, with \eqref{cons3}, the heuristic method selects randomly at most one robot of the set of robots whose receiving beams from the same RIS are overlapped together (condition to remove interference \cite{9681803}). Moreover, since the heuristic method always connects robots to BSs/RISs over the shortest paths, the allocation strategy from \eqref{cons9} and \eqref{cons10} cannot be applied, and the definitions of dummy variables from \eqref{add1} and \eqref{add2} are unnecessary. The heuristic checks the SINR QoS with \eqref{cons5} and \eqref{cons6}, the RIS allocation history with \eqref{cons7}, the available status of RISs with \eqref{cons8}. All inputs are summarized in Table \ref{tab:table3}. With $E=200$ elements, each RIS allows a maximum of $U=10$ robots being simultaneously served \cite{9681803}.

We randomly generate both the placements and the movement of robots in a factory setting.  Each robot moves five steps in the same direction before changing course to emulate more realistic movement patterns. We also generate a list of all RISs and BSs coverage areas in which a robot is in any given time slot. Each BS/RIS emits signals in conical beams with a beamwidth of $\theta$ degrees \cite{7820226,9822386}. Hence, we can calculate the footprint diameter $\phi$ of the conical beam from a BS/RIS to a robot with the distance $d$ as \cite{9386246}, $\phi = 2tan\left( \frac{\theta}{2} \right)d$. If any robots are within any undesired footprint, they are considered as receiving an interfering signal. The analytical results are run for $100$ random independent scenarios for each case, so we get the average values with confidence intervals of $95 \%$. 

We define the percentage of outages as the ratio of the total number of outages to the product of the total number of time slots and the number of robots. Also, the percentage of feasible solutions may happen due to an infeasible scenario or the proposed methods failing to find a feasible solution. 

\begin{table}[t!]
  \footnotesize
  \centering
    \vspace{0.2 cm}
  \caption{List of input parameters and binary variables of ILP program.}
  \label{tab:table3}
  \begin{tabular}{ll}
    \toprule
 \textbf{Input} & \textbf{Meaning}\\
    \midrule
    \footnotesize
    $|B|=2$ & Number of BSs. \\
$|I|=8$ & Number of RISs.\\
    $E=200$ & Number of RIS elements \cite{9681803}.\\
    $|N|=50$ & Number of time slots.\\
    $\theta=10^o$ & Beamwidth \cite{10214505}.\\
    $V=20 \cdot 10^6$ Hz & Bandwidth \cite{9306896}. \\
    $T=290$ Kelvin & Absolute temperature \cite{9306896}.\\
    $P_b=1$ mW & Power of BS.\\
    $f=28 \cdot 10^9$ Hz & Frequency \cite{9306896,7400962}.\\
  
\bottomrule
  \end{tabular}
\end{table}

\begin{figure*}[ht]
  \centering
\subfigure[Outage level]{\includegraphics[width=0.3\linewidth]{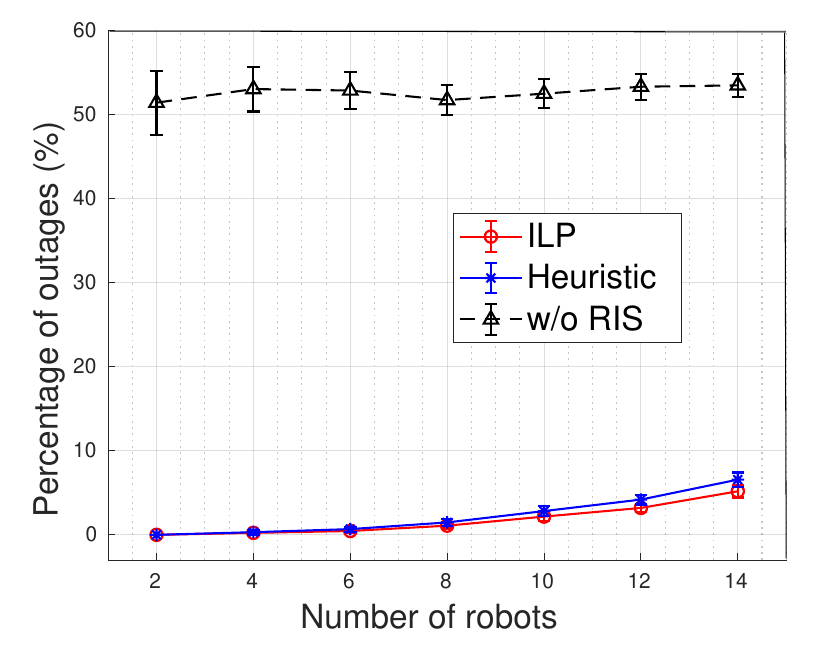} \label{outage_toward_R}}
\subfigure[Feasible solutions]{\includegraphics[width=0.3\linewidth]{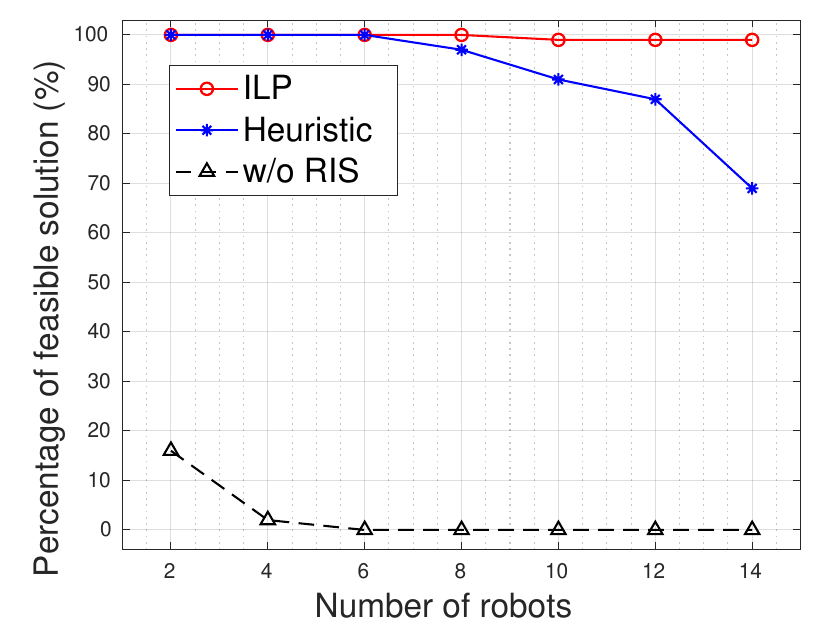} \label{feasible_toward_R}}
\subfigure[Time complexity]
{\includegraphics[width=0.3\linewidth]{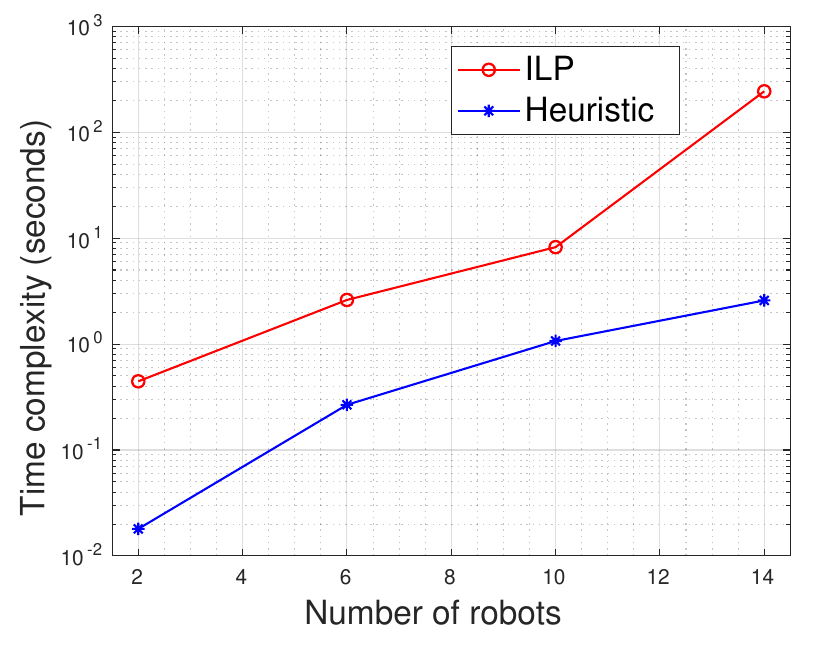} \label{time_toward_R}}
\caption{Outages, feasibility and complexity: $K_r \in [14,15]$ time slots, $\psi_r \in [9,10]$, $D=2$ time slots, and $U=2$ robots simultaneously served at each RIS.}\label{toward_R}
  \end{figure*} 
  
  Fig. \ref{toward_R} shows the percentage of outages and feasible solutions. We set $K_r=[14,15]$ time slots and threshold $\Psi_r=[9,10]$ for robots (averages $K_r=14.5$ and $\Psi_r=9.5$),  RIS configuration time $D=2$ time slots, and $U=2$ robots simultaneously served by a RIS. For comparison, a To analyze its a scenario without RIS (w/o RIS) is also analyzed. The percentage of outages in Fig. \ref{outage_toward_R} remains fairly constant w/o RIS with increasing number of robots, because this case only depends on the number of robots inside the BS coverage areas, which is around $(50,60) \%$. With RISs, both the ILP and heuristic methods significantly reduce the percentage of outages and enhance the percentage of feasible solutions. As expected, the ILP method outperforms the heuristic method, as shown in Fig. \ref{outage_toward_R} and in Fig. \ref{feasible_toward_R}, due to its optimal RIS allocation strategy. Regarding feasibility, the ILP method achieves up to $99 \%$ with $|R|=14$ robots, whereas the heuristic method achieves only under $70 \%$, as evidence of the hardness of this problem. Due to its competition for RIS connections, the percentage of outages and infeasible solutions increases with the number of robots. Finally, the time complexity of the algorithms is shown in Fig. \ref{time_toward_R}, whereas the exponential complexity of the ILP is shown for 14 robots. 

    \begin{figure*}[ht]
  \centering
\subfigure[$K_r=14;15$ time slots]{\includegraphics[width=4.2cm,height=4.2cm]{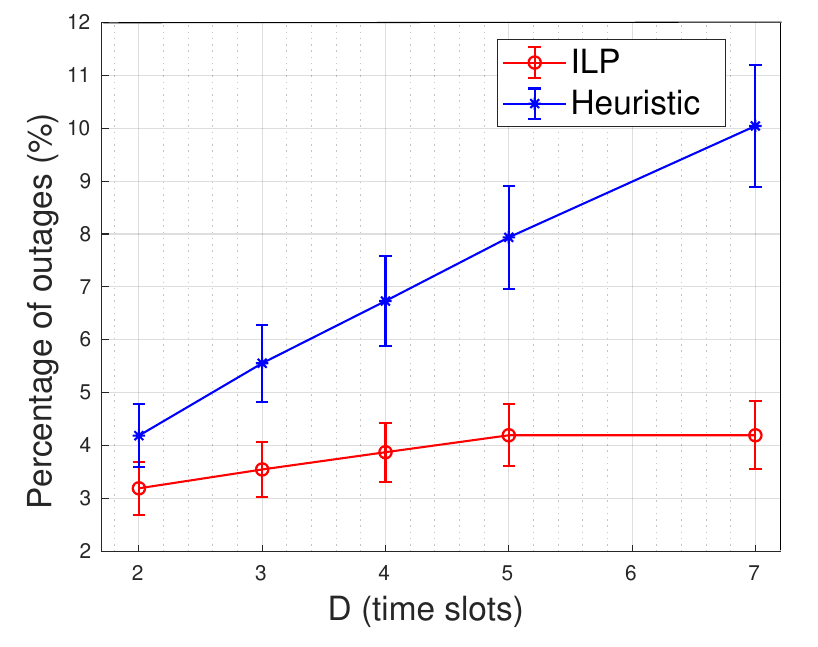} \label{outage_toward_D}}
\subfigure[$K_r=14;15$ time slots]{\includegraphics[width=4.2cm,height=4.2cm]{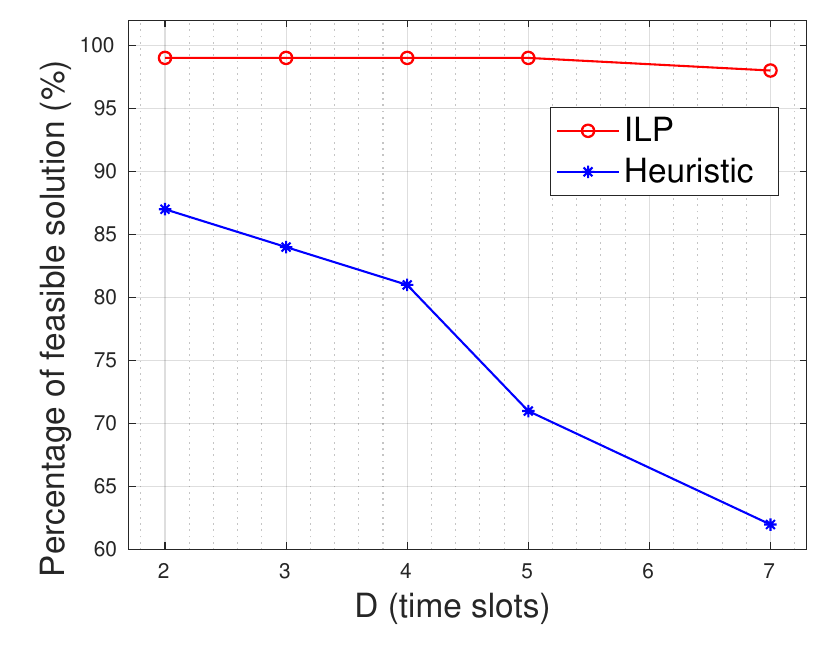} \label{feasible_toward_D}}
\subfigure[$D=2$ time slots]{\includegraphics[width=4.2cm,height=4.2cm]{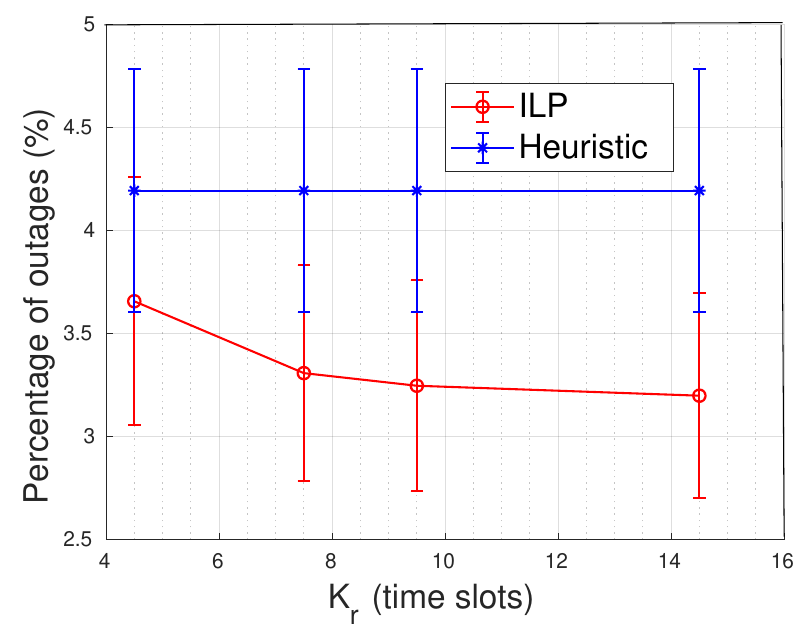} \label{outage_toward_Kr}}
\subfigure[$D=2$ time slots]{\includegraphics[width=4.2cm,height=4.2cm]{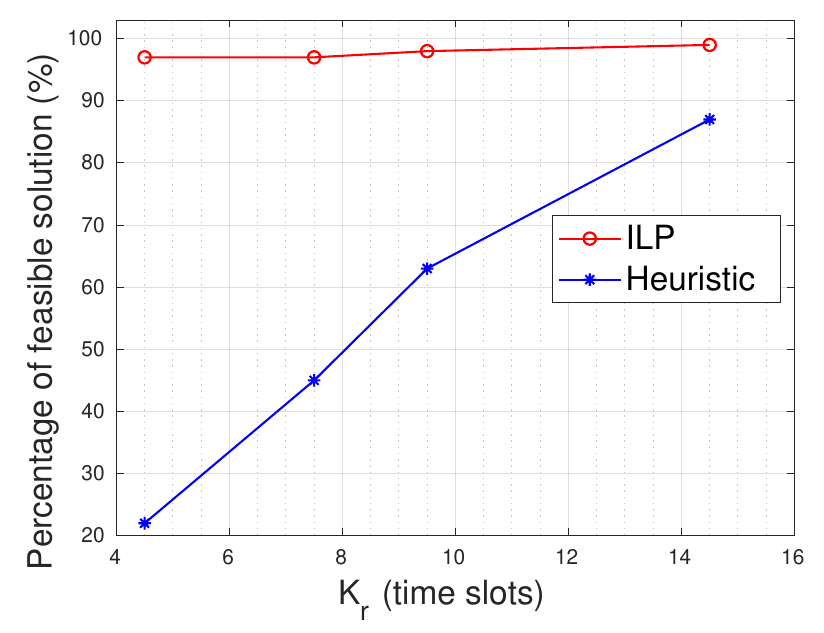} \label{feasible_toward_Kr}}
\caption{Percentage of outages and feasible solutions: $\psi_r \in [9,10]$, $U=2$ robots simultaneously served at each RIS, and $R=12$ robots.}\label{toward_D_Kr}
  \end{figure*} 

The impact of the reconfiguration time $D$ on the performance  is shown in Fig. \ref{outage_toward_D} and Fig. \ref{feasible_toward_D}, with $|R|=12$ robots. The percentage of outages and infeasible solutions increases with increasing $D$ due to increasing the unavailability period of an RIS after reallocation. The heuristic feasibility significantly decreases as $D$ increases. The impact of the maximum outage time $K_r$ (a QoS requirement) on the performance iis shown n Fig. \ref{outage_toward_Kr} and Fig. \ref{feasible_toward_Kr}, with $|R|=12$ robots. The average $K_r$ plotted are $K_r=[4.5;7.5;9.5;14.5]$. The percentage of outages of the heuristic remains constant for all $K_r$, because it does not consider the allocation strategy for robots (constraints (\ref{cons9}) and \eqref{cons10}). Hence, the heuristic is independent on the values of $K_r$ and consequently shows less feasibility. For optimizations, with increasing $K_r$, we can find solutions with a lower percentage of outages (Fig. \ref{outage_toward_Kr}) and infeasible solutions (Fig. \ref{feasible_toward_Kr}) because QoS requirements of $K_r$ are low. Again, the heuristic cannot cope with the majority of scenarios, especially for lower $K_r$ values.

      \begin{figure*}[ht]
  \centering
\subfigure[$U=2$ robots and $R=12$ robots]{\includegraphics[width=4.2cm,height=4.2cm]{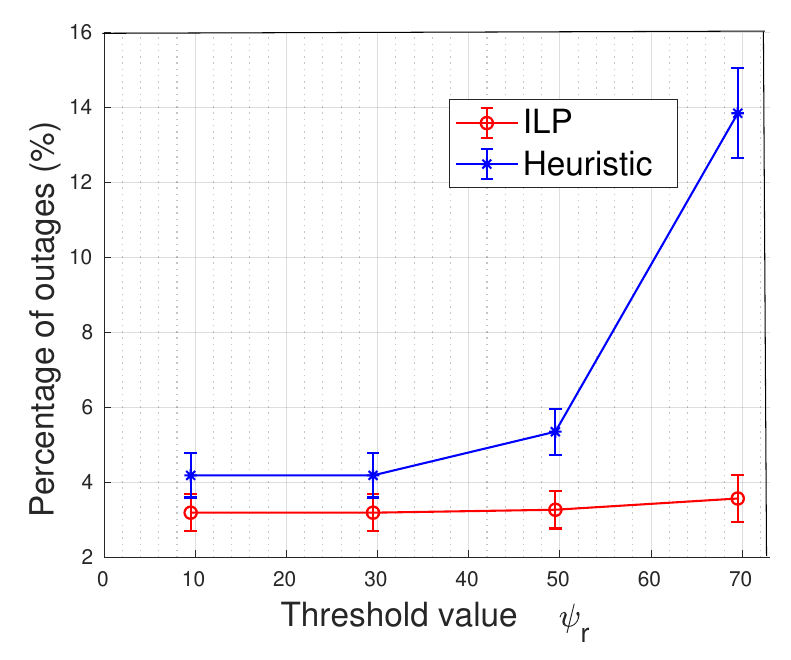} \label{outage_toward_psi}}
\subfigure[$U=2$ robots and $R=12$ robots]{\includegraphics[width=4.2cm,height=4.2cm]{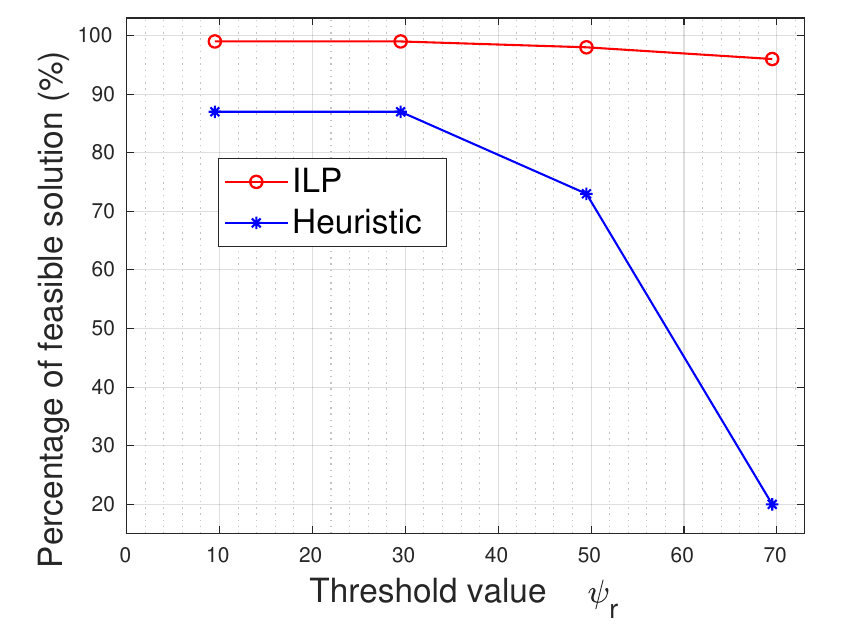} \label{feasible_toward_psi}}
\subfigure[$\psi_r = 9;10$ and $R=14$ robots]{\includegraphics[width=4.2cm,height=4.2cm]{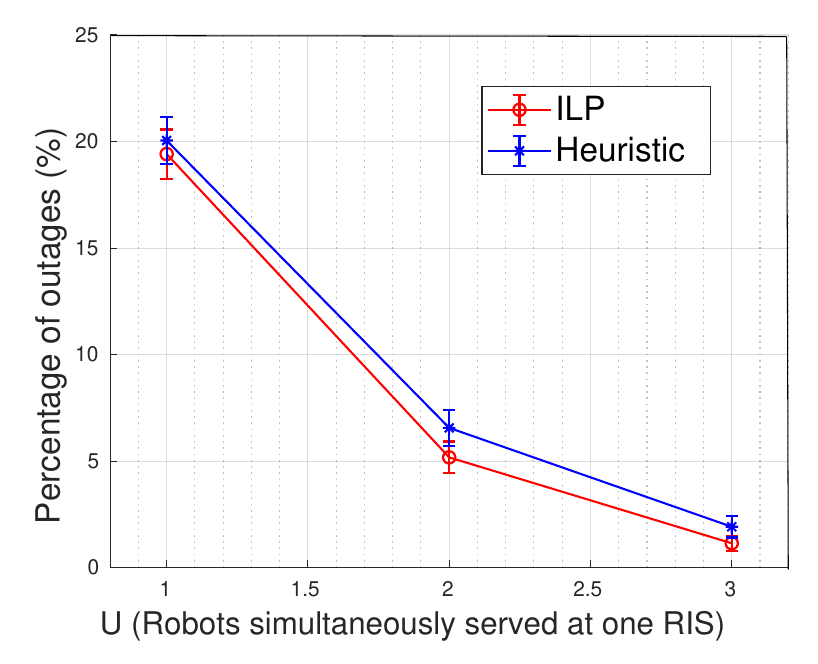} \label{outage_toward_U}}
\subfigure[$\psi_r = 9;10$ and $R=14$ robots]{\includegraphics[width=4.2cm,height=4.2cm]{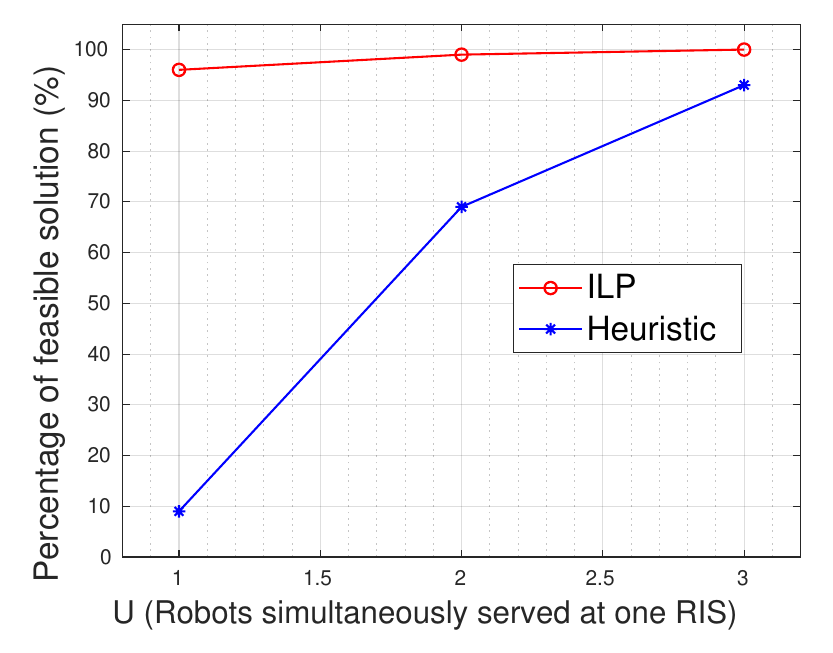} \label{feasible_toward_U}}
\caption{Percentage of outages and feasible solutions: $D=2$ time slots and $K_r \in [14,15]$.}\label{toward_psi_U}
  \end{figure*}

  In Fig. \ref{outage_toward_psi} and Fig. \ref{feasible_toward_psi}, we analyze the impact of the SINR threshold (a QoS requirement) on performance, whereas $|R|=12$ robots. The average values $\Psi_r$ plotted are $\Psi_r=[9.5;29.5;49.5;69.5]$. It is harder to find more feasible solutions with increasing $\Psi_r$ in Fig. \ref{feasible_toward_psi} because the SINR requirements are more stringent, resulting in a higher percentage of outages (Fig. \ref{outage_toward_psi}).  Due to optimal interference avoiding and RIS allocation strategies, ILP solution does not show major difference with increasing $\Psi_r$. Finally, we analyze the performance impact of $U$ robots served simultaneously at each RIS in Fig. \ref{outage_toward_U} and Fig. \ref{feasible_toward_U}, with $|R|=14$. We find fewer outages and infeasible solutions with increasing $U$ because each RIS allows more robots to schedule transmission simultaneously. Thus, it becomes easier to find more feasible solutions. We stop at $U=3$ because the ILP achieves up to $100\%$ feasibility (Fig. \ref{feasible_toward_U}). This also means that, for the scenario with 14 robots, we could employ RIS devices with fewer reflecting elements, which is more cost effective. 
  
  \section{Conclusion} \label{conclu}
We formulated a novel combinatorial optimization problem, of which solution can optimally maintain connectivity in smart factory with robots, by optimally allocating RISs to robots. Our model exploited the characteristics of nulling interference from RISs by tuning RIS reflection coefficients. We also considered  connection outages due to insufficient transmission quality service, and especially signal-to-interference-plus-noise ratio (SINR), which was novel. Numerical results showed the benefits of RISs in reducing the number of link outages. We also showed that interference-avoiding strategies were necessary to optimize the performance. The study revealed further research directions, towards multi-hop RIS-assisted mmWave networks as well as towards more effective heuristics in controlled environment of future smart factories. 

\bibliographystyle{IEEEtran}

\bibliography{all_refs}

\end{document}